\documentclass[useAMS,usenatbib]{mnras}

\newcommand{\includepgf}[1]{\IfFileExists{#1}{\input{#1}}{}}

\usepackage{xcolor}

\usepackage{graphicx}
\usepackage{pgf}
\usepackage{float}
\usepackage{import}
\usepackage{subcaption}
\usepackage{microtype}

\usepackage{amsmath}
\usepackage{amssymb}
\usepackage{bm}

\usepackage{cleveref}
\crefname{equation}{Equation}{Equations}
\crefname{figure}{Figure}{Figures}
\crefname{section}{Section}{Sections}
\crefname{table}{Table}{Tables}

\title[]{Varying Driver Velocity Fields in Photospheric MHD Wave Simulations}
\author[A. J. Leonard, S. J. Mumford, V. Fedun, R. Erd\'elyi]{
    A. J. Leonard$^{1}$,
    S. J. Mumford$^{1}$,
    V. Fedun$^{2}$ and
    R. Erd\'elyi$^{1}$.\\
    $^{1}$Solar Physics \& Space Plasma Research Centre (SP$^{2}$RC), School of Mathematics and Statistics,\\\ \ The University of Sheffield, Hicks Building, Hounsfield Road, Sheffield, S3 7RH U.K.\\
    $^{2}$Department of Automatic Control and Systems Engineering, The University of Sheffield, Mappin Street, Sheffield, S1 3JD, U.K.\\ }
\begin{document}
\label{firstpage}
\pagerange{\pageref{firstpage}--\pageref{lastpage}}
\maketitle

\begin{abstract}
Torsional motions are ubiquitous in the solar atmosphere.
In this work, we perform 3D numerical simulations which mimic a vortex-type photospheric driver with a Gaussian spatial profile.
This driver is implemented to excite MHD waves in an axially symmetric, 3D magnetic flux tube embedded in a realistic solar atmosphere.
The Gaussian width of the driver is varied and the resulting perturbations are compared.
Velocity vectors were decomposed into parallel, perpendicular and azimuthal components with respect to pre-defined magnetic flux surfaces.
These components correspond broadly to the fast, slow and Alfv\'en modes, respectively.
From these velocities the corresponding wave energy fluxes are calculated, allowing us to estimate the contribution of each mode to the energy flux.
For the narrowest driver ($0.15$ Mm) the parallel component accounts for $\sim 55 - 65\%$ of the flux.
This contribution increases smoothly with driver width up to nearly $90\%$ for the widest driver ($0.35$ Mm).
The relative importance of the perpendicular and azimuthal components decrease at similar rates.
The azimuthal energy flux varied between $\sim 35\%$ for the narrowest driver and $< 10\%$ for the widest one.
Similarly, the perpendicular flux was $\sim 25 - 10\%$.
We also demonstrate that the fast mode corresponds to the sausage wave in our simulations.
Our results therefore show that the fast sausage wave is easily excited by this driver and that it carries the majority of the energy transported.
For this vortex-type driver the Alfv\'en wave does not contribute a significant amount of energy.
\end{abstract}


\section{Introduction} \label{sec:intro}
Magnetohydrodynamic (MHD) waves are ubiquitous in the solar atmosphere and it is considered likely by many that they contribute to solar atmospheric heating by transporting energy from the photosphere up through the lower solar atmosphere and into the low corona.
There have been numerous observations in various magnetic structures of each of the MHD wave modes - fast, slow and Alfv\'en.
The fast mode, in particular, is frequently seen having been observed in sunspots \citep[\emph{e.g.}][]{dorotovic_standing_2014}, pores \citep[\emph{e.g.}][]{morton_observations_2012, dorotovic_standing_2014, freij_detection_2014} and other magnetic structures in the chromosphere \citep{morton_observations_2012}.
\cite{dorotovic_standing_2014} also observed the slow mode.
Alfv\'en waves have been observed in a group of bright points by \cite{jess_alfven_2009}, and \cite{mcintosh_alfvenic_2011} claim to have detected them in the corona.
For reviews of the wide range and variety of wave observations see e.g. \cite{nakariakov_coronal_2005, bogdan_observational_2006, zaqarashvili_oscillations_2009, wang_standing_2011, mathioudakis_alfven_2013, 2013ApJ...769...44S, de_moortel_transverse_2016}.

Torsional motions have great potential to excite Alfv\'en waves, the favourite candidate for energy transport in solar MHD (see \emph{e.g.} \cite{mathioudakis_alfven_2013} for a review of Alfv\'en wave observations and theory).
Therefore torsional motions have been searched for and have been successfully observed at \emph{e.g.} intergranular lanes in the form of resolution-limited small-scale vortices \citep[\emph{e.g.}][]{bonet_convectively_2008, wedemeyer-bohm_small-scale_2009, bonet_sunrise/imax_2010}.
It is widely accepted that these vortices form due to turbulent convection \citep[\emph{e.g.}][]{shelyag_photospheric_2011, wedemeyer-bohm_magnetic_2012, kitiashvili_ubiquitous_2013}.

Given the ubiquity of these vortex motions in the photosphere, it is important to understand how the waves they excite contribute to the heating of the lower solar atmosphere.
To this end, several three-dimensional simulations have been performed by \citep[\emph{e.g.}][]{fedun_mhd_2011, vigeesh_three-dimensional_2012, wedemeyer-bohm_magnetic_2012, mumford_generation_2015}.
These studies implemented torsional motions at the base of a realistic magnetic flux tube and analysed the resulting perturbations.
In each case it was found that such a driver excites fast and slow magnetoacoustic waves and the Alfv\'en wave, and that in all but one case the sausage and kink modes were both present.
\cite{vigeesh_three-dimensional_2012} and \cite{mumford_generation_2015} also quantified the energy flux of waves produced by torsional motions and found that the azimuthal components of the waves made a greater contribution to the flux than the perpendicular or parallel components.

The present work is a continuation of the work of \cite{mumford_generation_2015}, which investigates the effects of varying driver parameters on the wave motions stimulated by those drivers in the low solar atmosphere.
In this case, we now implement a spiral velocity driver and investigate how varying that driver's width scales the wave energy transport from the driver into the lower solar atmosphere.
Since a range of vortex sizes are observed, we wish to investigate whether this variation causes different waves, as this information will have implications for atmospheric heating.
We also outline a new way to unambiguously demonstrate the presence of sausage and kink modes (whether slow or fast, depending on the equilibrium conditions) in our simulations by calculating the displacement of the magnetic flux surface from its original position.

For this study we use the Sheffield Advanced Code \citep[SAC;][]{shelyag_magnetohydrodynamic_2008}, which is built on the basis of the Versatile Advection Code \citep[VAC;][]{toth_general_1996}.
SAC separates variables into background and perturbed components, allowing the simulation of highly gravitationally stratified media such as the solar atmosphere.

This paper is structured as follows: in  Section \ref{sec:simsetup} we describe the background atmosphere and the properties of the photospheric drivers employed in the simulations; in Section \ref{sec:sims} we describe the simulation parameters and the analysis method; in Section \ref{sec:results} we present the results of the simulations and the analysis; in Section \ref{sec:conclusions} we discuss these results and present our conclusions.

\section{Background atmosphere and photospheric drivers} \label{sec:simsetup}

\begin{figure}
    \centering
    \includegraphics[width=0.95\columnwidth]{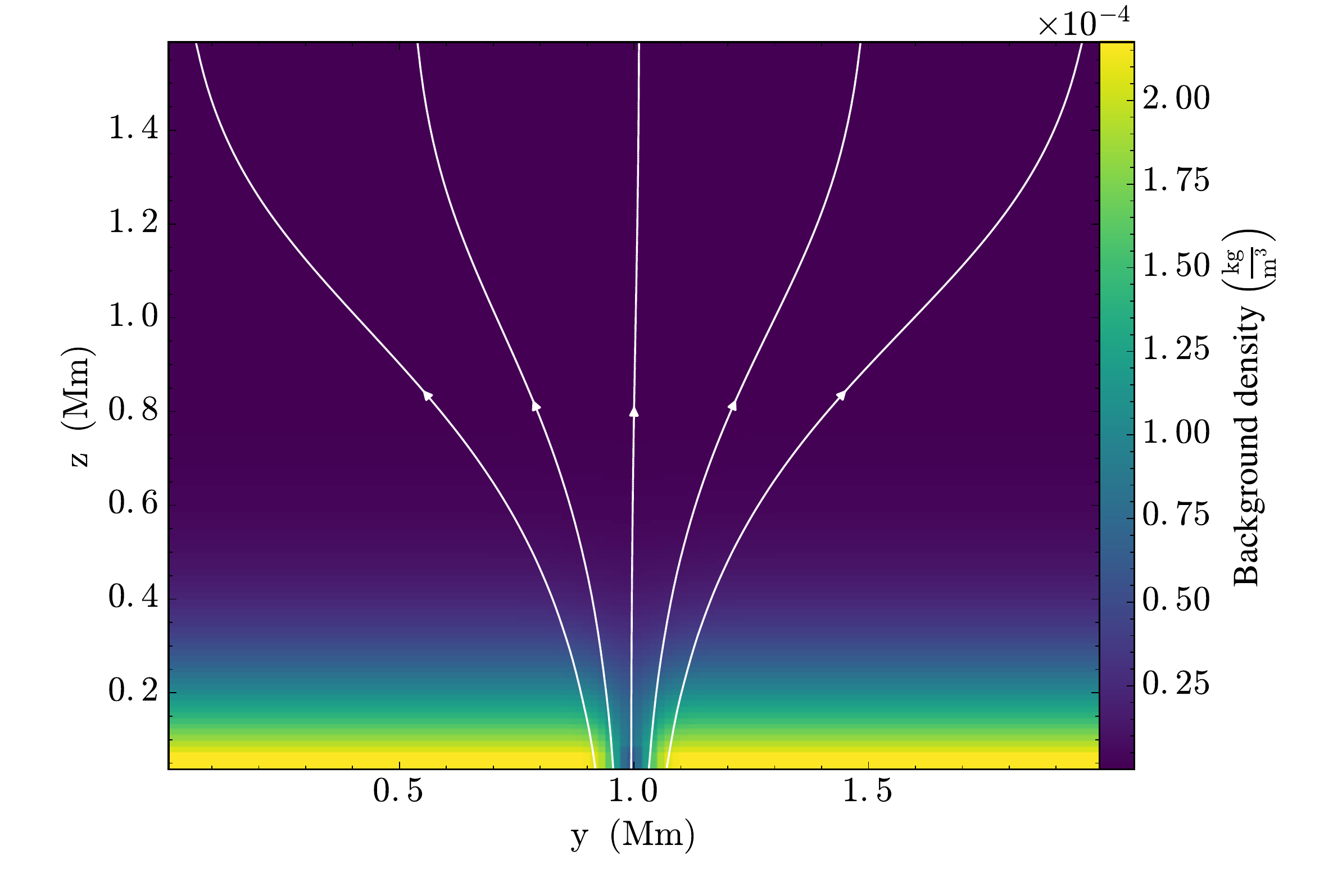}
    \caption{Initial distribution of background density plotted on a two-dimensional slice through the centre of the domain in the $x$ direction.
    Overplotted white streamlines correspond to the magnetic field lines and indicate the shape of the magnetic flux tube in the background atmosphere.}
    \label{fig:ini}
\end{figure}

\begin{figure}
    \centering
    \includegraphics[width=0.95\columnwidth]{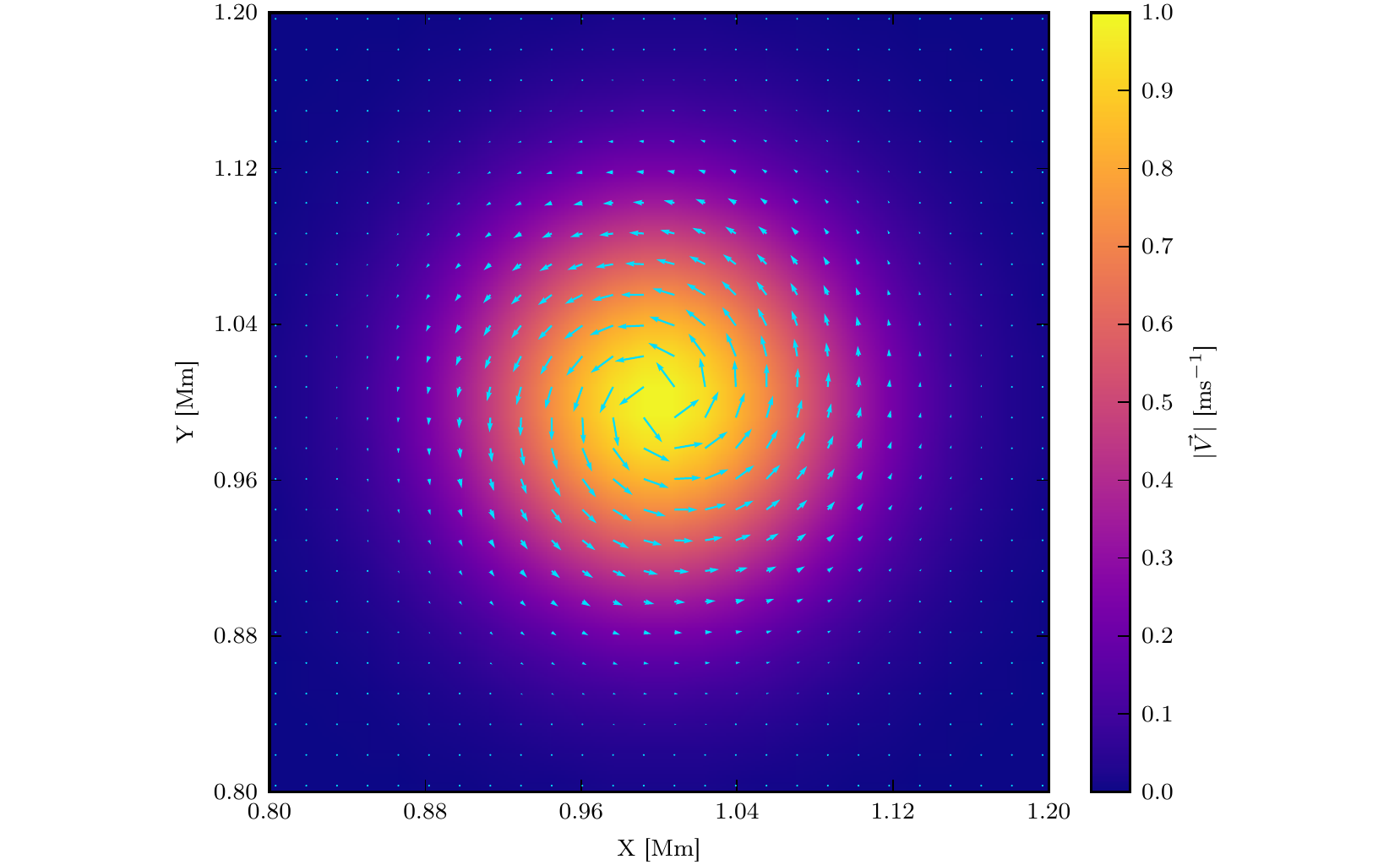}
    \caption{Normalised horizontal profile of the velocity field generated by the implemented spiral driver.
    Colour-coding indicates the magnitude of the velocity and the cyan arrows indicate the direction.
    This velocity profile is multiplied by the magnitude of the driver (see text) and varies with both height and time.}
    \label{fig:slog-profile}
\end{figure}

Here, we use a three dimensional background atmosphere (Figure \ref{fig:ini}) based on the VAL IIIC model \citep{vernazza_structure_1981}, and implement an axisymmetric magnetic flux tube modelled based on the self-similar approach \citep{schluter_internal_1958, deinzer_magneto-hydrostatic_1965, schussler_dynamical_2005, fedun_numerical_2011, gent_magnetohydrostatic_2013}.
The full-width at half-maximum of the magnetic flux tube (FWHM) was approximately $90$ km in the photosphere.
The footpoint of the magnetic flux tube was centred at $x, y, z = (1.0, 1.0, 0.0)$ Mm, and had a magnetic field strength of $143.6$ mT (see \cref{fig:ini}).
For more details on this initial configuration, see \cite{mumford_generation_2015} - who use the same background atmosphere - and references therein.

In each simulation, perturbations to the background atmosphere are driven by introducing a velocity field in the horizontal plane close to the footpoint of the flux tube.
These drivers are intended to mimic different kinds of velocity fields that may be found in the photosphere, as a result of granulation.
This paper uses the same logarithmic spiral velocity fields as \cite{mumford_generation_2015}, to study the excitation of torsional waves in the atmosphere.
The logarithmic spiral shape is based on observations by \cite{bonet_convectively_2008} and others of vortex flows in intergranular lanes (see Section \ref{sec:intro}).
The logarithmic spiral driver is centred on the point $x, y, z = (1.0, 1.0, 0.1)$ Mm, the spatial extent of which is determined by a Gaussian profile in each direction.
The velocity at a given point and at time $t$ is described by:
\begin{subequations}
\begin{align}
	v_x &=   A \frac{\cos(\theta + \phi)}{\sqrt{x^2 + y^2}}\ G(x, y, z) \sin \left(2\pi \frac{t}{P}\right),\\
	v_y &= - A \frac{\sin(\theta + \phi)}{\sqrt{x^2 + y^2}}\ G(x, y, z) \sin \left(2\pi \frac{t}{P}\right),
\end{align}
\label{eq:slog}
\end{subequations}
where
\begin{equation*}
  G(x, y, z) = \exp\left(-\frac{z^2}{\Delta z^2} - \frac{x^2}{\Delta x^2} - \frac{y^2}{\Delta y^2}\right)
\end{equation*}
is the Gaussian profile with width $\Delta x$, $\Delta y$ and $\Delta z$ in the $x$-, $y$- and $z$-directions, respectively.
$A$ is the driver amplitude, $P$ is the driver period, $\theta = \tan^{-1}(y/x)$ is the angle around the flux tube axis and $\phi = \tan^{-1}(1/0.15)$ determines the expansion of the spiral.
In a series of numerical simulations, \citep[][Chapter 6]{mumford_2016_48888} found that the period of such a swirly motion has only a relatively minor effect on the results of this kind of study, with the contribution from the Alfv\'en mode, for instance, varying by less than $20\%$.
The value of the period for this study was therefore mostly chosen so that a few periods would fit into the run-time of the simulation, and was set to $90$ s.
The choice of the expansion parameter, $0.15$, was also largely arbitrary and was selected to allow a few rotations of the spiral within the driver.

We use five values for the horizontal Gaussian width of the driver, $\Delta x = \Delta y$, as indicated in Table \ref{tab:period-amp}.
This range of parameter values was chosen to correspond to the range of major axes found by \cite{sanchez_almeida_bright_2004} for a sample of 126 magnetic bright points (MBPs) observed in intergranular lanes, and are consistent with Bonet's observations that these vortexes have sizes of $\leq 0.5$ Mm.
Each driver had the same vertical width, $\Delta z = 0.05$ Mm.

All drivers were designed to supply the same total amount of energy, $E_T$, into the simulation.
To ensure this, the amplitude of the driver was adjusted according to the width of the driver.
The exact relation is
\begin{equation}
  E_T = \frac{n P V A^2}{4} \sum_{x, y, z} \rho(x, y, z) G^2(x, y, z) = const.
\end{equation}
In the above, $V$ is the volume of the computational domain, $\rho$ is the density, and $n$ is the integer number of periods in the simulation run-time.
The actual amplitudes corresponding to the widths implemented in the simulations are listed in Table \ref{tab:period-amp}.

\begin{table}
\centering
 \begin{tabular}{ccc}
 Width (Mm) & Width (FWHM) & Amplitude (ms$^{-1}$) 	\\ \hline
 $0.15$     & $1.67$           & $10.221$   	\\[2ex]
 $0.20$     & $2.22$           & $7.465$  		\\[2ex]
 $0.25$     & $2.78$           & $5.894$      \\[2ex]
 $0.30$     & $3.33$           & $4.875$     \\[2ex]
 $0.35$     & $3.89$           & $4.159$     \\[2ex]
 \end{tabular}
 \caption{Driver width $\Delta x = \Delta y$ and corresponding driver amplitude values used to ensure the same input of total kinetic energy to each simulation.
   The middle column indicates the ratio of the width of the driver to the FWHM of the flux tube.}
 \label{tab:period-amp}
\end{table}

\section{Simulations} \label{sec:sims}

\begin{figure*}
    \centering

    \begin{subfigure}[b]{\columnwidth}
        \includegraphics[width=0.95\columnwidth]{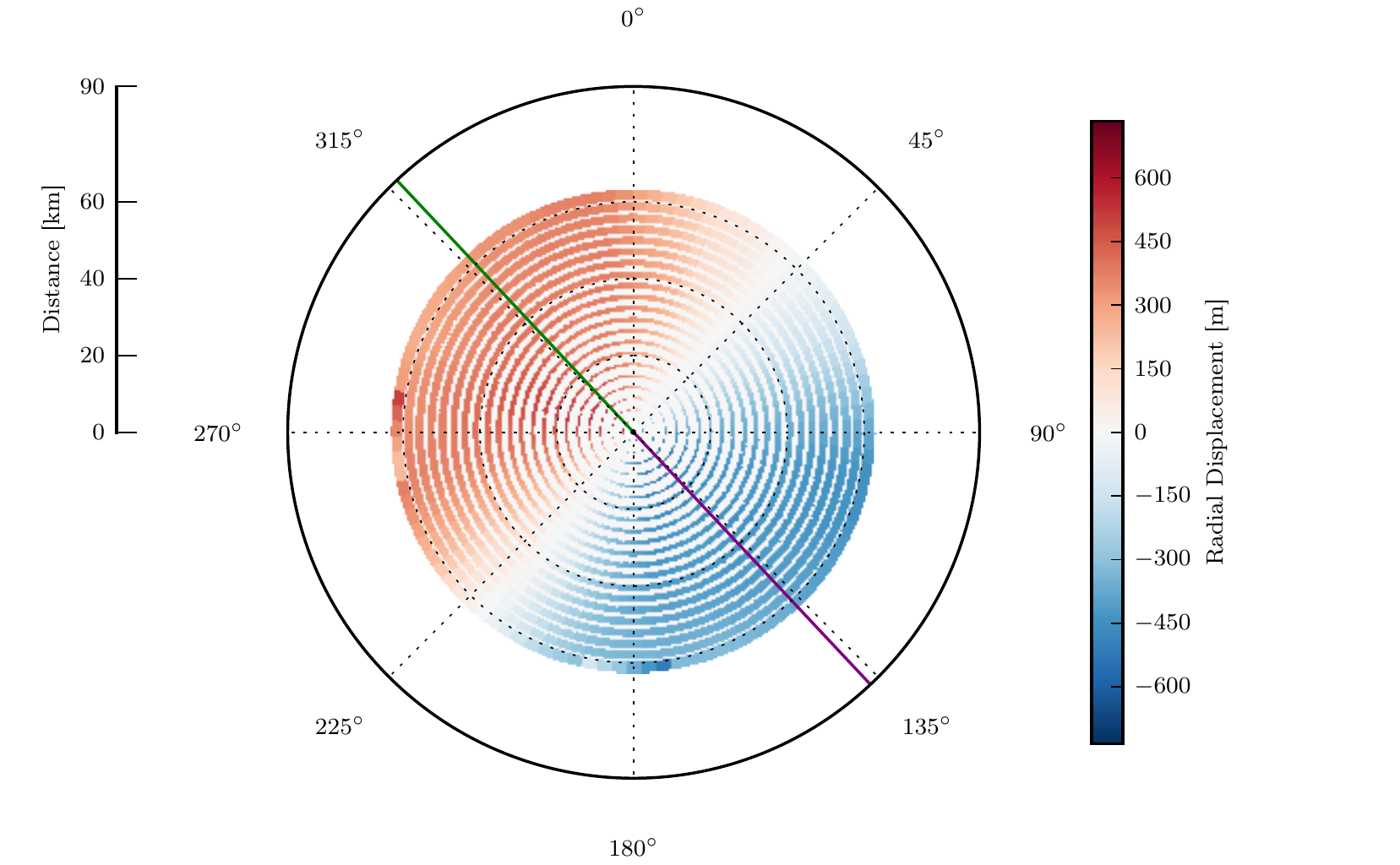}
        \caption{Height: 187.5 km}
        \label{fig:rdisplacement-hslice-Slog-p90-0-10--0-15-h015-t210}
    \end{subfigure}
    \begin{subfigure}[b]{\columnwidth}
        \includegraphics[width=0.95\columnwidth]{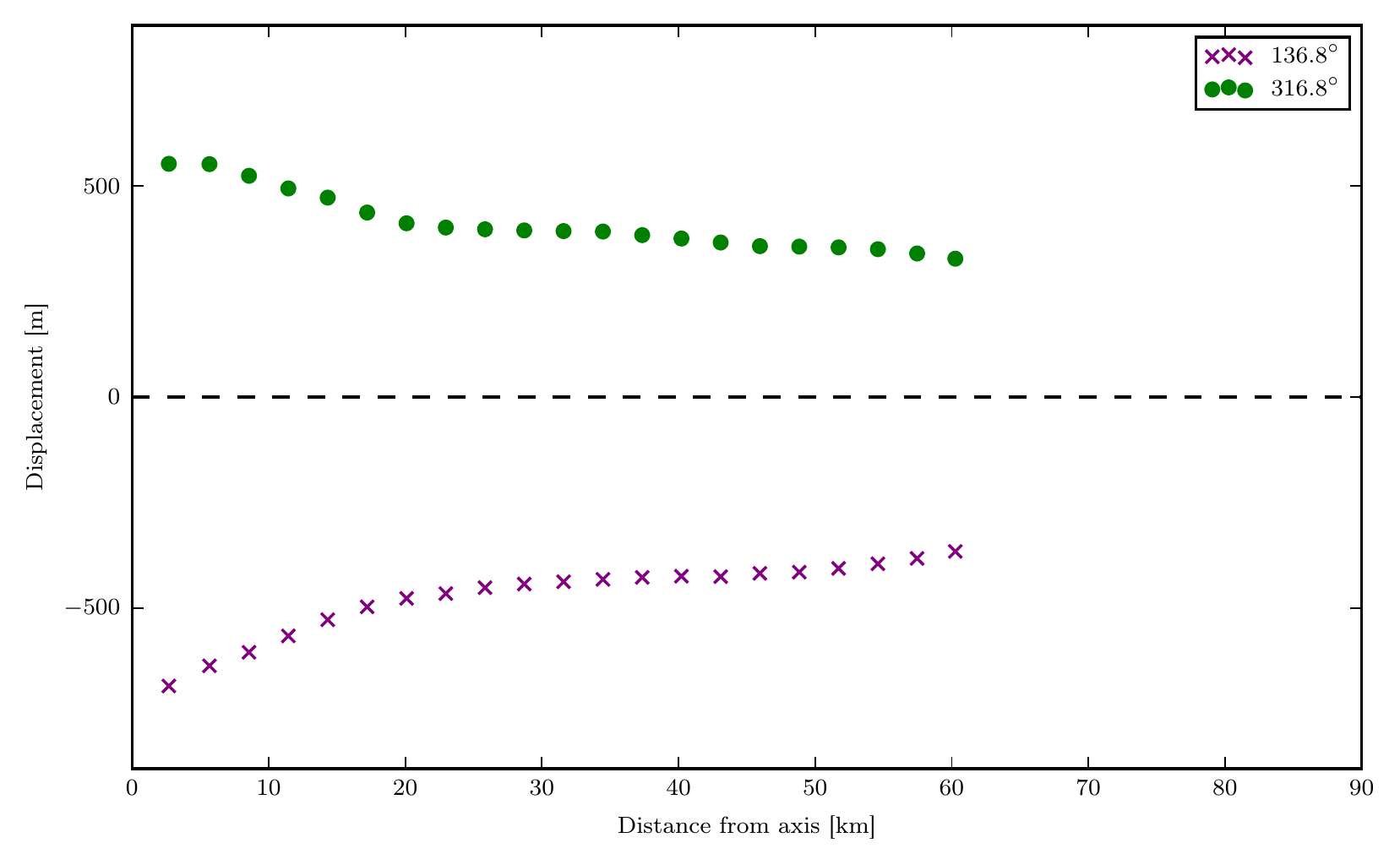}
        \caption{Height: 187.5 km}
        \label{fig:rdisplacement-hline-Slog-p90-0-10--0-15-h015-t210}
    \end{subfigure}

    \begin{subfigure}[b]{\columnwidth}
        \includegraphics[width=0.95\columnwidth]{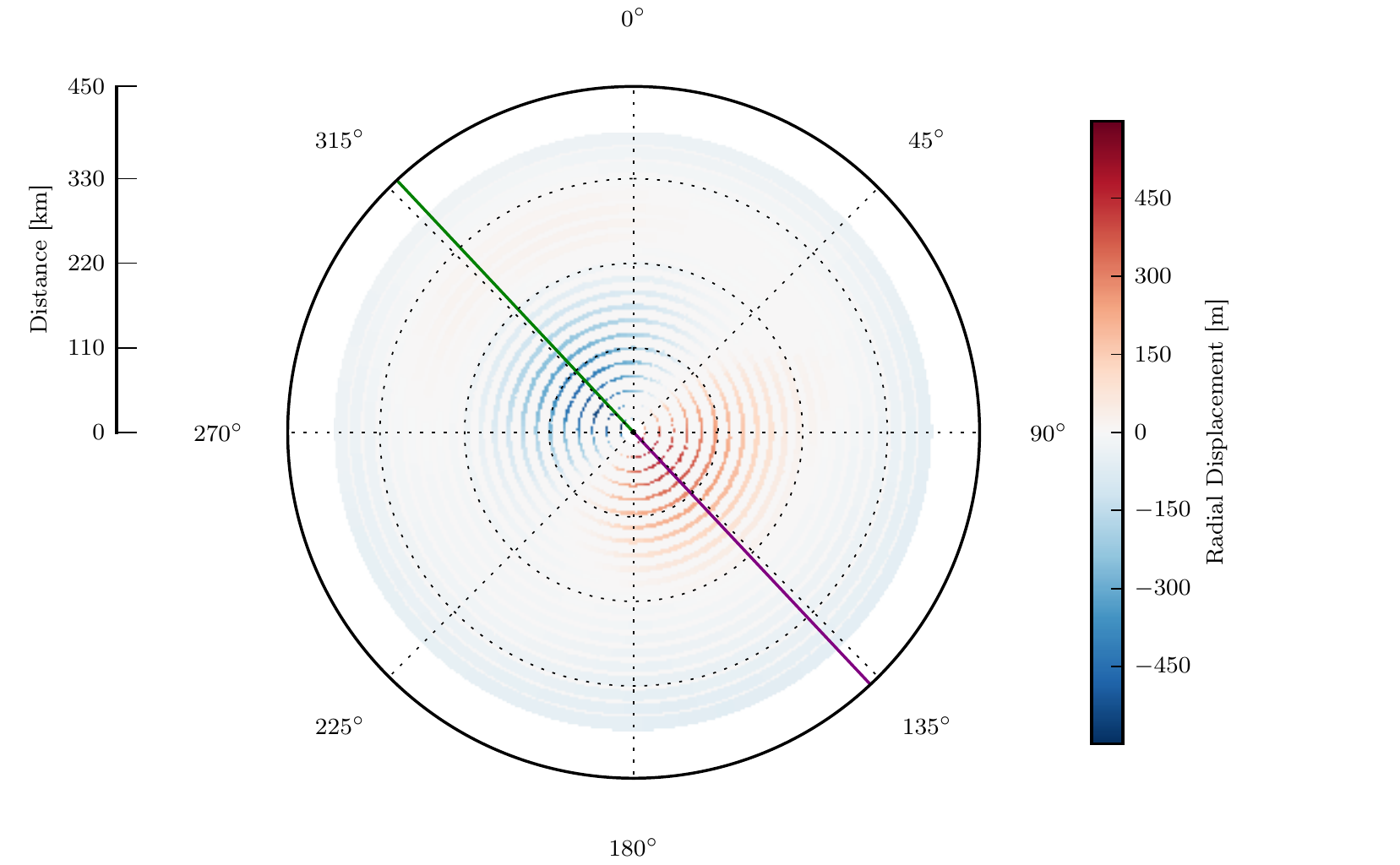}
        \caption{Height: 812.5 km}
        \label{fig:rdisplacement-hslice-Slog-p90-0-10--0-15-h065-t210}
    \end{subfigure}
    \begin{subfigure}[b]{\columnwidth}
        \includegraphics[width=0.95\columnwidth]{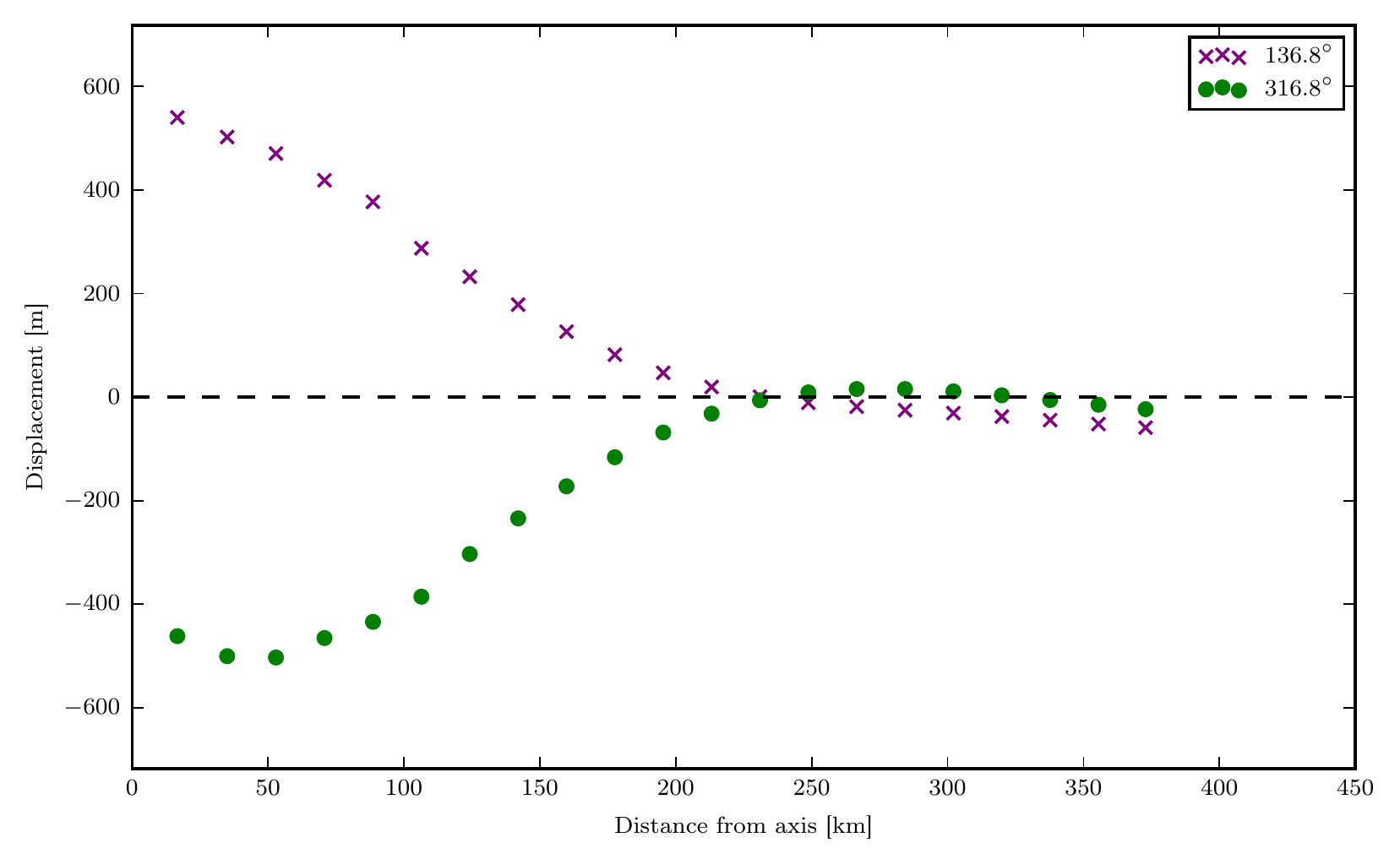}
        \caption{Height: 812.5 km}
        \label{fig:rdisplacement-hline-Slog-p90-0-10--0-15-h065-t210}
    \end{subfigure}

    \begin{subfigure}[b]{\columnwidth}
        \includegraphics[width=0.95\columnwidth]{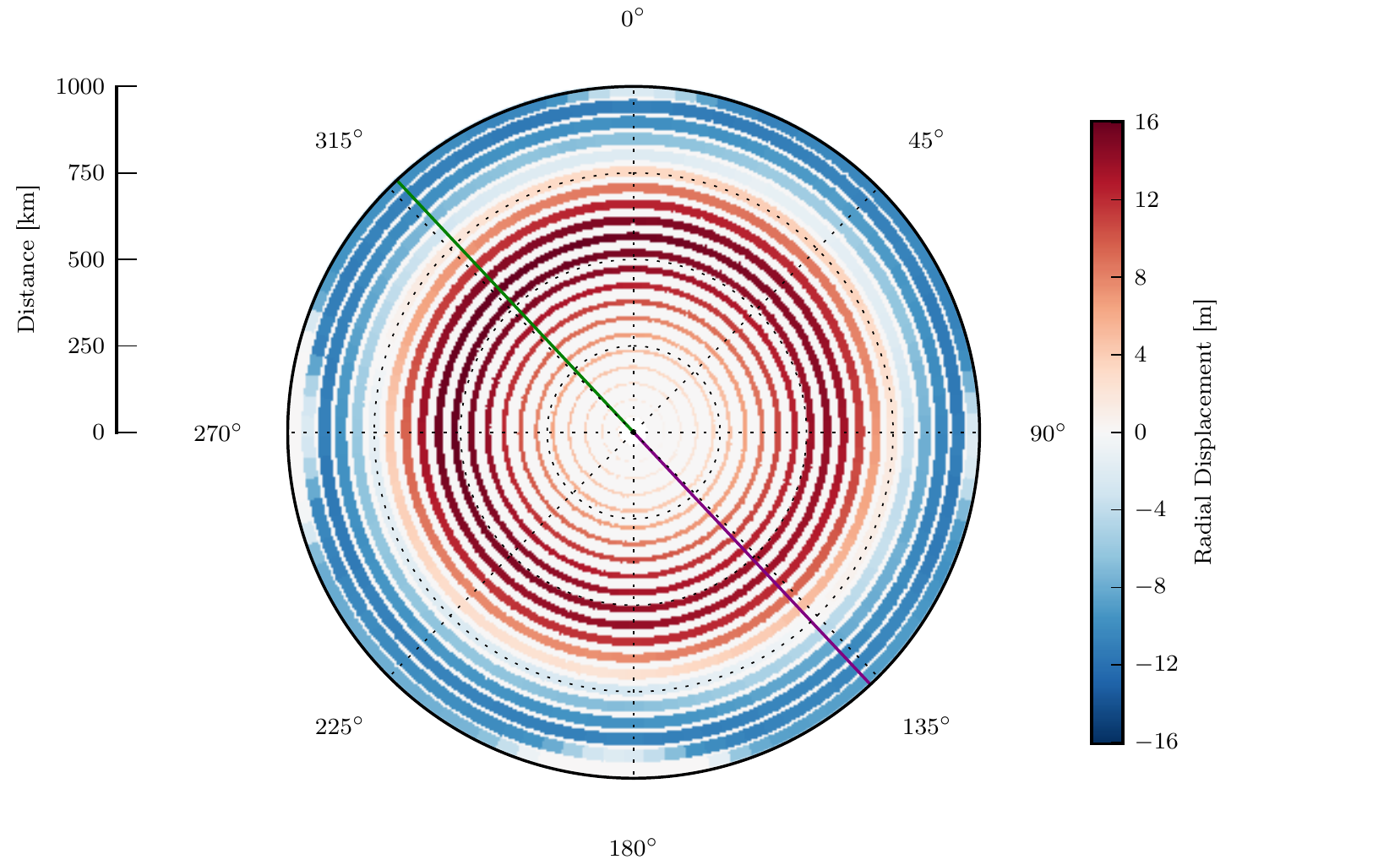}
        \caption{Height: 1437.5 km}
        \label{fig:rdisplacement-hslice-Slog-p90-0-10--0-15-h115-t210}
    \end{subfigure}
    \begin{subfigure}[b]{\columnwidth}
        \includegraphics[width=0.95\columnwidth]{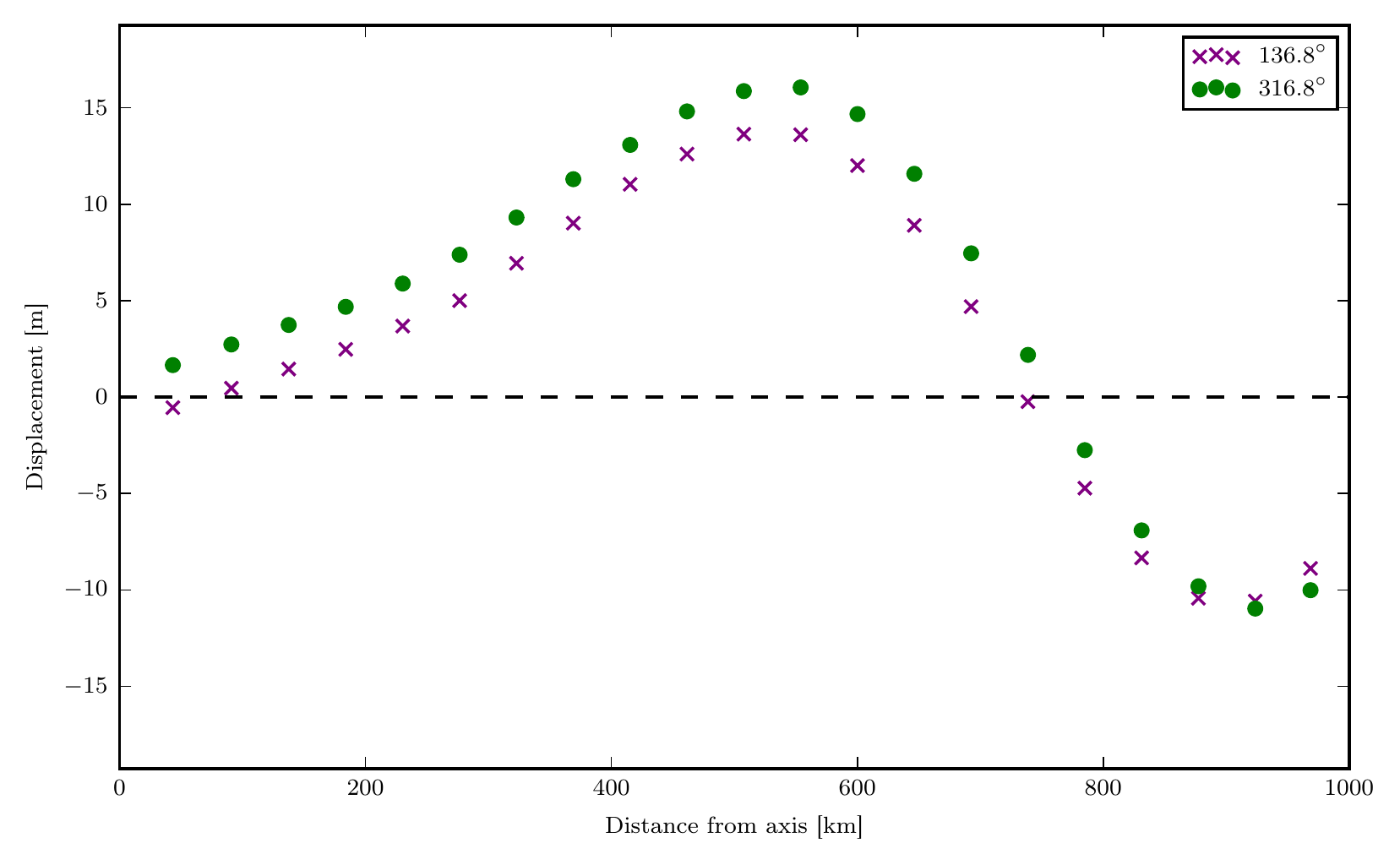}
        \caption{Height: 1437.5 km}
        \label{fig:rdisplacement-hline-Slog-p90-0-10--0-15-h115-t210}
    \end{subfigure}
    \caption{Radial displacement of points located on several flux surfaces throughout the domain at time $t = 214.5$ s.
    Note that the radial extent on each of these plots is different due to the expansion of the flux tube at greater heights in the domain.
    Left column: polar plots showing motion of the points away from (red) or towards (blue) the flux tube axis.
    The radial and azimuthal axes indicate the location of each point with respect to the axis.
    Right column: displacement of points located at $136.8^\circ$ (purple crosses) and  $316.8^\circ$ (green circles), indicated on the polar plots by radial lines of the same colour.
    We can see see in these plots evidence of the kink mode low in the domain and the sausage mode near the top.
    In between we see both waves, with the kink confined close to the flux tube axis and the sausage seen closer to the edges.
}
    \label{fig:hslices}
  \end{figure*}

The simulation domain ranged from $0.0$ Mm to $2.0$ Mm in the $x$- and $y$-directions, and from $0.0$ Mm to $1.6$ Mm in the $z$-direction, with a mesh size of $128$, resulting in $128^3$ grid cells.
The boundaries of the domain were set to the `continuous' setting in SAC (\emph{i.e.:} the gradient of each variable was zero across the boundaries).
Each simulation was run for $270$ s of simulation time, equal to three full driver periods.
This amount of time is approximately equal to the lifetime of vortex flows observed in the photosphere by \cite{bonet_convectively_2008}.
We can therefore be confident that the flux tube would reasonably remain stable within the runtime of the simulation.

\subsection{Velocity vector decomposition}
A flux surface is constructed by selecting seed points on a circle near the top of the domain centred on the flux tube axis.
Field lines are traced down through the domain from those seed points to the bottom of the domain using the method of \cite{mumford_generation_2015}.
These field lines then enclose a constant amount of magnetic flux at any given height and thus describe the surface of a flux tube.
We refer to these field lines and flux surfaces by the radius of the circle of seed points, as a fraction of the maximum radius possible in the domain (64 grid cells).
These field lines are retraced from these advected seeds at each time-step and new flux surfaces are calculated.

This treatment allows us to separate the velocities into components which are locally parallel to the direction of the magnetic field, perpendicular to the magnetic flux surface and azimuthal around the flux tube.
These components correspond broadly to the fast and slow MHD waves and the Alfv\'en wave, respectively.

Of course, with this interpretation of wave modes, one has to bear in mind the local value of the plasma-${\beta}$, where ${\beta}$ is the ratio of kinetic to magnetic pressure.
Given the fairly weak magnetic field of our flux tube, plasma-$\beta$ is large ($> 1$) everywhere in the simulation domain except for a small region at the top of the domain close to the flux tube axis.
Therefore the slow mode propagates mainly along magnetic field lines with the local Alfv\'en speed, $v_A$.
The fast mode is allowed to propagate in any direction.
Along field lines the fast mode travels with the sound speed $c_s$, while in the direction perpendicular to the magnetic field it propagates with the phase speed $v_p = \sqrt{c_s^2 + v_A^2}$.

\subsection{Velocity and flux calculation}
Following \cite{mumford_generation_2015}, we interpolate the decomposed velocity vectors onto a single field line in order to study how waves propagate along this field line throughout the simulation.
In addition to the velocity components, we define the wave energy flux using the following equation \citep{leroy_derivation_1985, bogdan_waves_2003, mumford_generation_2015}:
\begin{equation}
  \bm{F}_{wave} = \tilde p_k \bm{v} + \frac{1}{\mu_0} (\bm{B}_b \cdot \bm{\tilde{B}}) \bm{v} - \frac{1}{\mu_0} (\bm{v} \cdot \bm{\tilde{B}}) \bm{B}_b,
  \label{eq:waveflux}
\end{equation}
where $\tilde p_k$ is the kinetic pressure perturbation,
\begin{equation}
  \tilde p_k = (\gamma - 1) (\tilde e - \frac{\rho \bm{v}^2}{2} - \frac{\bm{B}_b \bm{\tilde{B}}}{\mu_0} - \frac{\bm{\tilde{B}}^2}{2 \mu_0}).
  \label{eq:kinpress}
\end{equation}
Here, $\bm{v}$ is the velocity, $\bm{B}$ is the magnetic field, $e$ is the total energy density, $\mu_0$ is the permeability of free space and $\gamma$ is the adiabatic index of the plasma.
Background and perturbed components of quantities are indicated by a subscript $b$ and a tilde, respectively.
Once calculated, the energy flux vector can be decomposed into its parallel, perpendicular and azimuthal components in the same way as the velocity vector.

\subsection{Flux surface displacement}
We calculate the distance at a number of angular positions around the axis by finding the intersection of the flux surfaces with lines through the axis at those angles.
These distances are calculated for a number of heights in the domain and for each time-step.
The values are then subtracted from the original distances at $t = 0$ s, giving the radial displacement with respect to the original positions of the surfaces.
This displacement describes the distortion of the flux tube, which allows us to determine whether waves are sausage or kink by comparing the direction of displacement on opposite sides of the flux tube.

The sausage wave, which distorts the flux tube in the same direction (\emph{i.e.}: towards or away from the axis) at all angles, will manifest as the displacement of any two points having the same sign.
Conversely, the kink mode moves the flux tube towards the axis on one side and away from it on the opposite side, resulting in negative and positive displacement of points on those sides, respectively.

Torsional motions could be detected by inspecting azimuthal velocity, $v_{\theta}$, using the same method, but doing so is not trivial due to technical limitations - this will be addressed in a later work.
This part of the analysis is therefore intended only to determine the presence of sausage and kink motions.

\section{Simulation results} \label{sec:results}

\subsection{Wave mode identification}

\begin{figure}
    \centering
    \includegraphics[width=0.95\columnwidth]{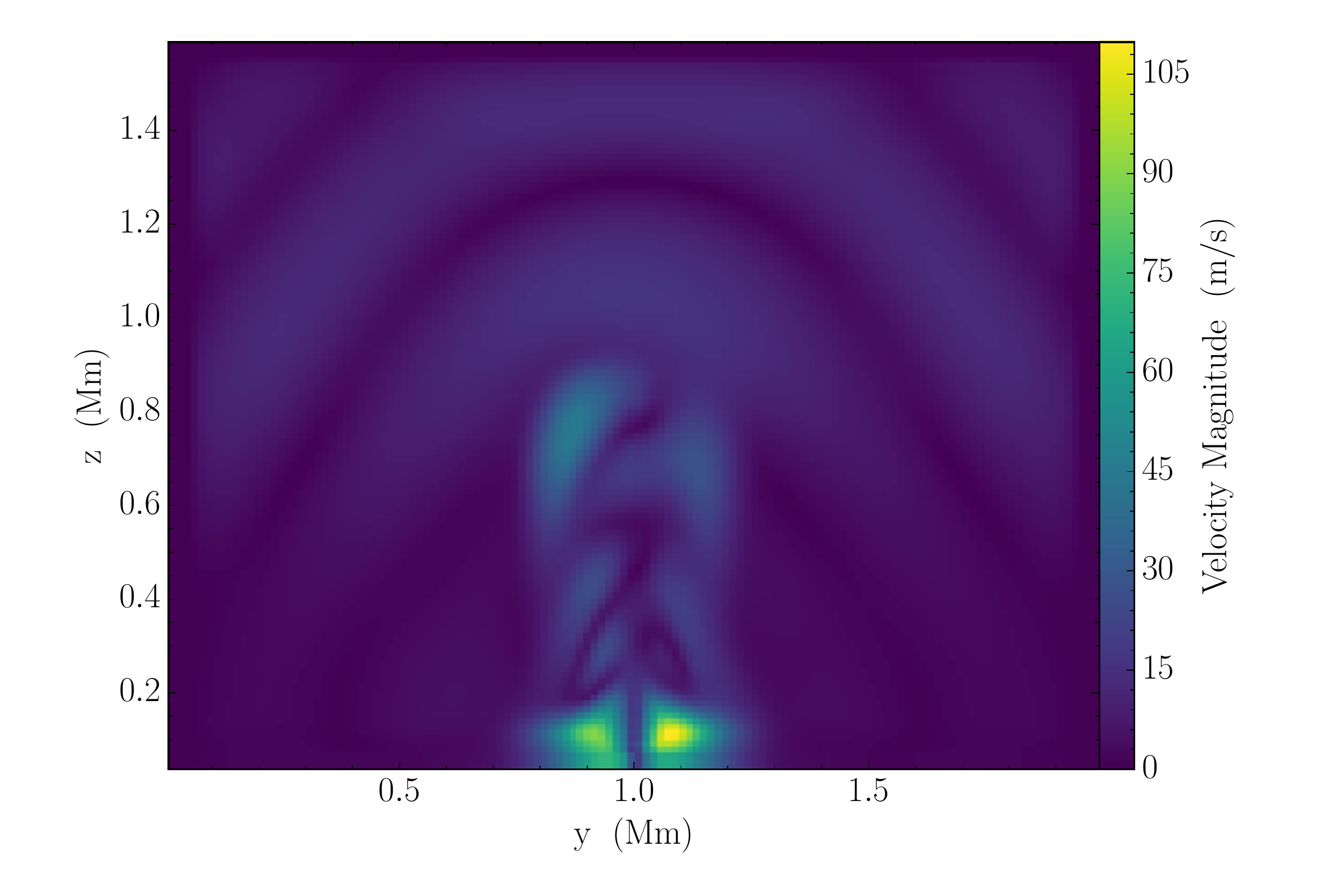}
    \caption{Magnitude of velocity on a vertical slice through the domain for the narrowest driver at time $t = {}$ s.
    Two sets of wave fronts are clearly visible, one propagating almost isotropically and the other closely following the axis of the flux tube.The isotropic wave has also reached the top of the domain by this time, whereas the other wave has only reached a little over half-way up, indicating that the latter is travelling much more slowly.
}
    \label{fig:velmag}
  \end{figure}

\begin{figure*}
    \centering

    \begin{subfigure}[b]{\columnwidth}
        \includegraphics[width=0.95\columnwidth]{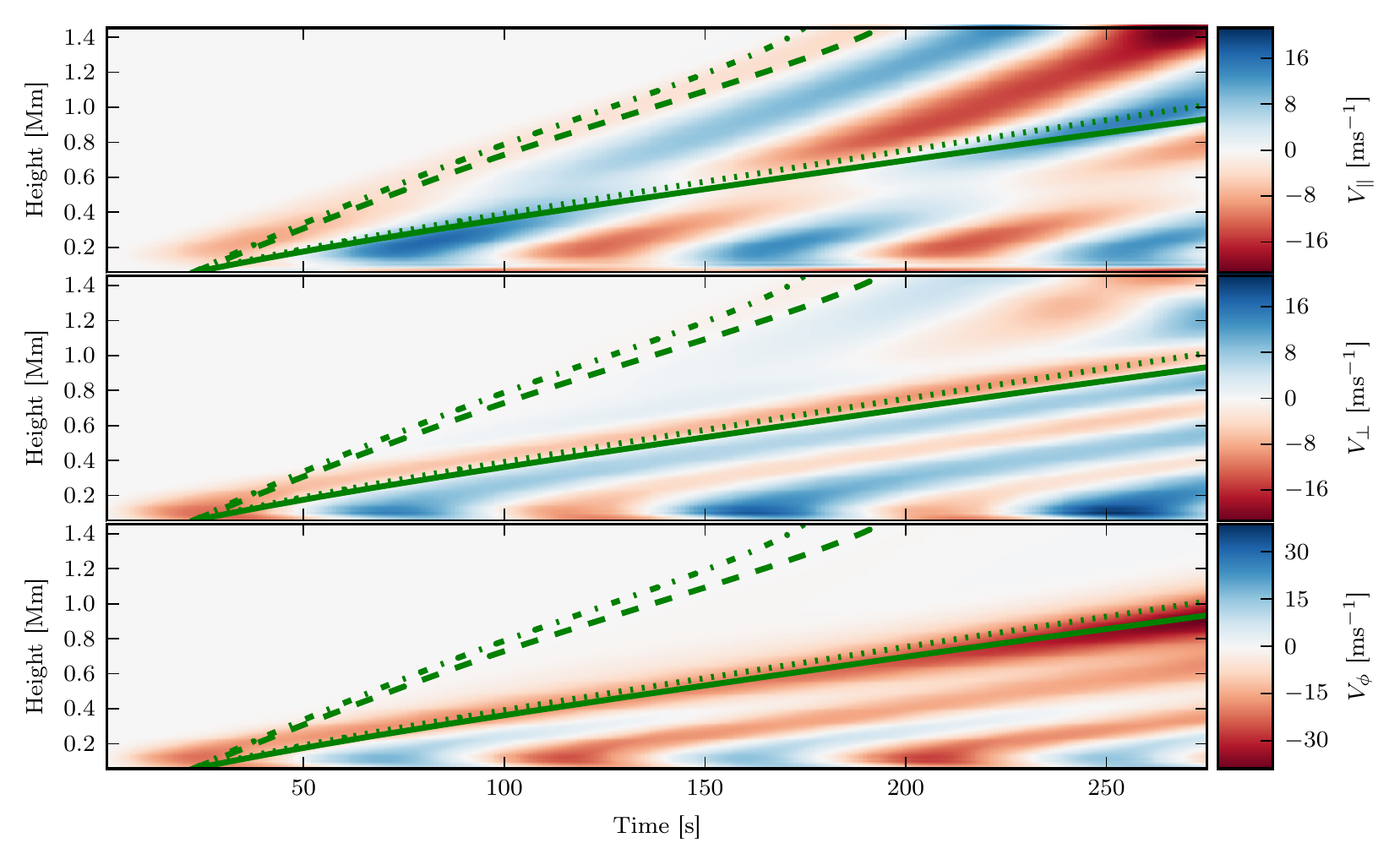}
        \caption{Driver width: 0.15 Mm}
        \label{fig:TD-vel-r30-w0-15}
    \end{subfigure}
    \begin{subfigure}[b]{\columnwidth}
        \includegraphics[width=0.95\columnwidth]{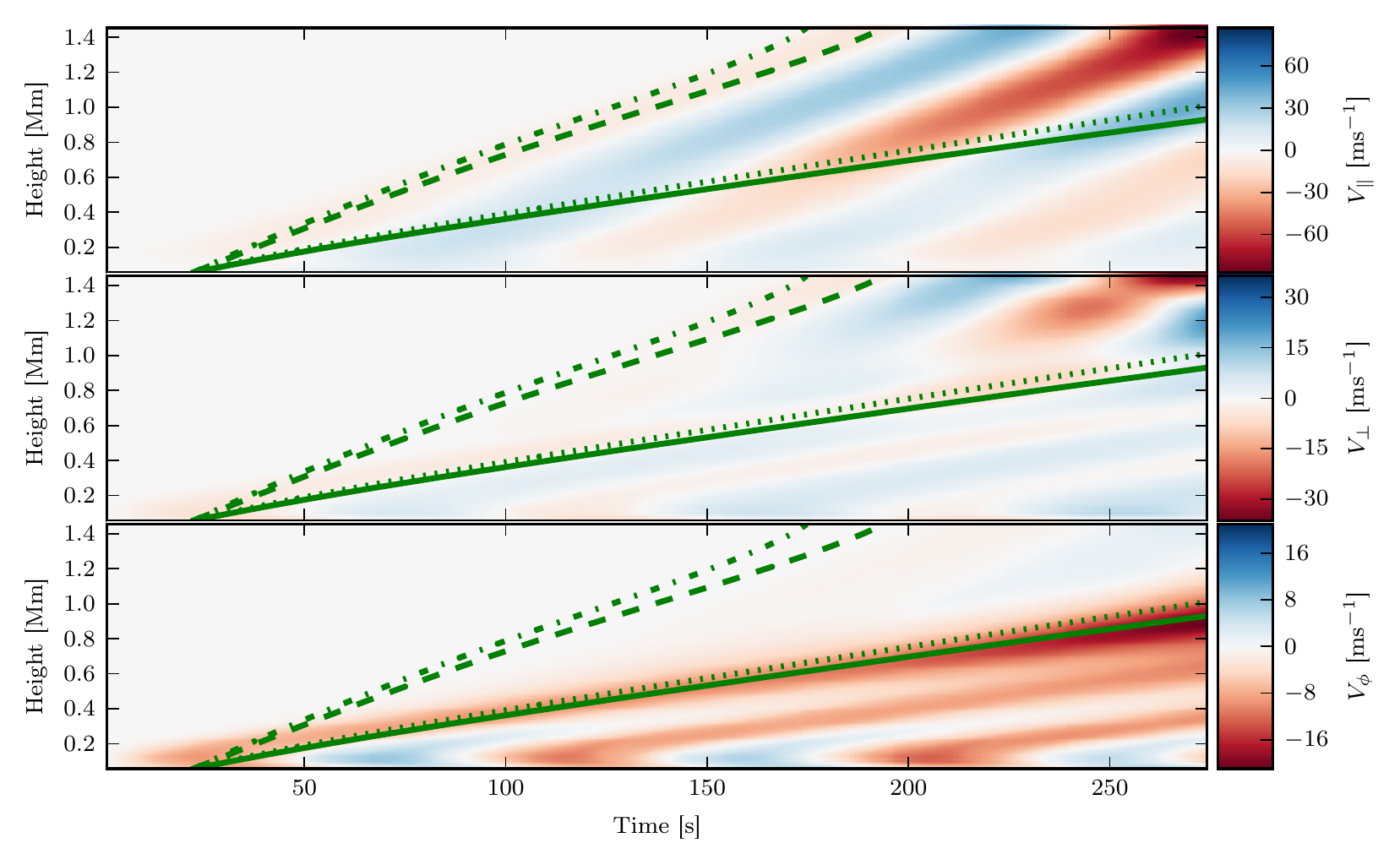}
        \caption{Driver width: 0.35 Mm}
        \label{fig:TD-vel-r30-w0-35}
    \end{subfigure}
    \caption{Time-distance diagrams of decomposed velocity components along a field line at $r = 0.469$ for the narrowest (left) and widest (right) drivers.
Each subplot shows the component of velocity parallel to the magnetic field ($v_{\parallel}$, top), the component perpendicular to the magnetic flux surface ($v_{\perp}$, middle) and the azimuthal component ($v_{\theta}$, bottom).
Overplotted lines indicate the fast ($v_f$, dot-dashed line), slow ($v_t$, solid line), sound ($c_s$, dashed line), and Alfv\'en ($v_A$, dotted line) speeds along the field line.}
    \label{fig:vel-td}
\end{figure*}

\begin{figure*}
    \centering

    \begin{subfigure}[b]{\columnwidth}
        \includegraphics[width=0.95\columnwidth]{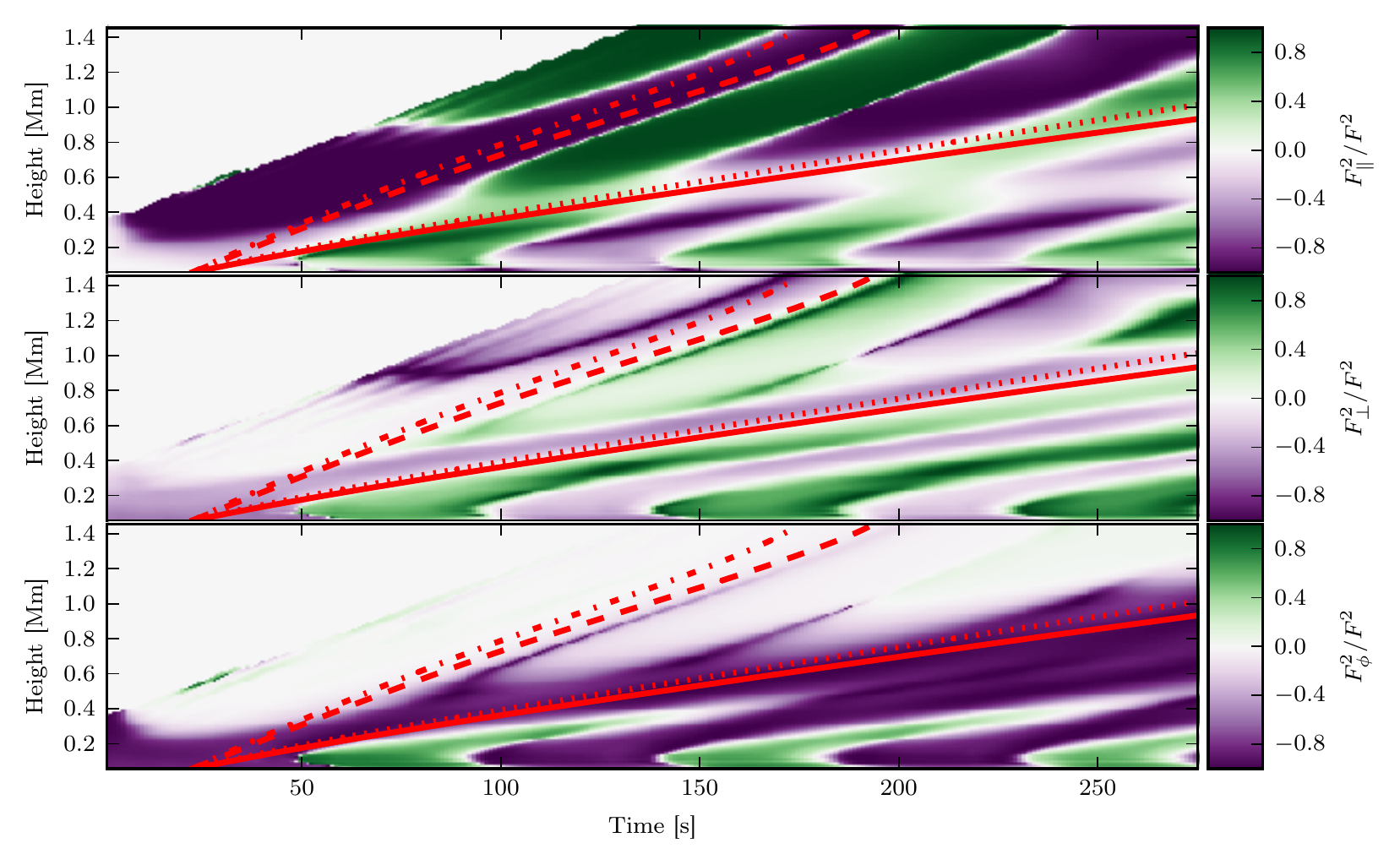}
        \caption{Driver width: 0.15 Mm}
        \label{fig:TD-flux-r30-w0-15}
    \end{subfigure}
    \begin{subfigure}[b]{\columnwidth}
        \includegraphics[width=0.95\columnwidth]{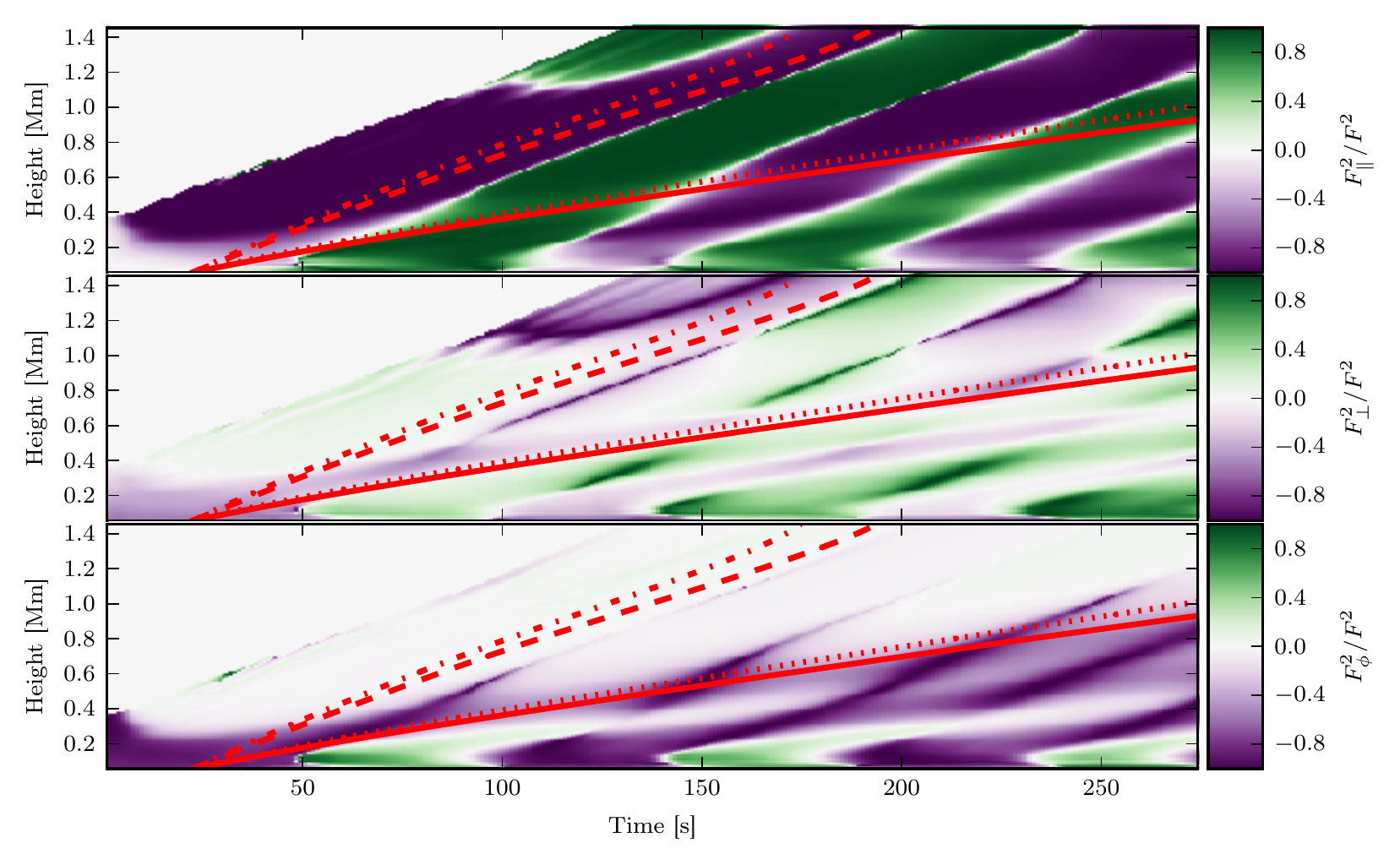}
        \caption{Driver width: 0.35 Mm}
        \label{fig:TD-flux-r30-w0-35}
    \end{subfigure}
    \caption{Same plots as Figure \ref{fig:vel-td} for the fractional square wave flux along the field line.}
    \label{fig:flux-td}
\end{figure*}

\begin{figure}
    \centering
    \includegraphics[width=0.95\columnwidth]{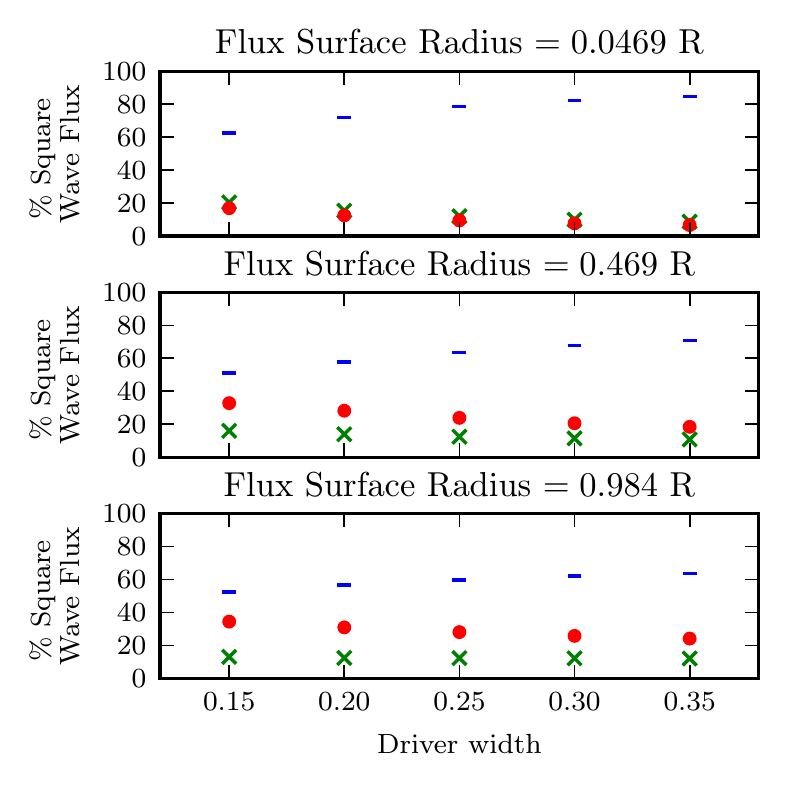}
    \caption{Average percentage square wave flux against spiral velocity driver width.
In each subplot, parallel, perpendicular and azimuthal flux components are indicated by blue dashes, green crosses and red circles, respectively.
The top, middle and bottom panels plot the flux for field lines at $r = 0.0469$, $r = 0.469$ and $r = 0.984$, respectively.
}
    \label{fig:fluxcompare-width}
\end{figure}

\begin{figure}
    \centering
    \includegraphics[width=0.95\columnwidth]{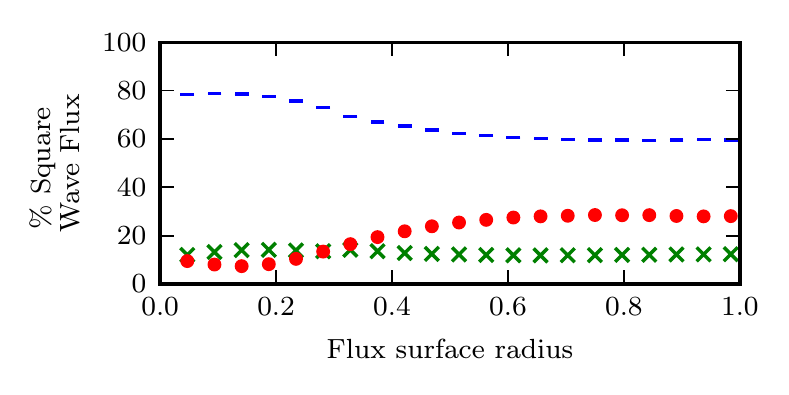}
    \caption{Average percentage square wave flux against flux surface radius for the driver with width $0.25$ Mm ($2.78$ FWHM).
Parallel, perpendicular and azimuthal flux components are indicated by blue dashes, green crosses and red circles, respectively.
}
    \label{fig:fluxcompare-radius}
\end{figure}

The displacement of flux surfaces with respect to their original positions was calculated as described above for twenty magnetic flux surfaces throughout the domain.
The radii of the seed point circles for these flux surfaces were equally spaced between $r = 0.047$ and $r = 0.984$ at intervals of $3$ grid cells, and each seed point was initially located $10$ grid cells below the top of the domain at $z = 1.475$ Mm.

Figure \ref{fig:hslices} shows the displacement of these flux surfaces at three different heights in the domain.
The plots in the left-hand column of this figure show the displacement of each point from its original position.
In these plots the radial and azimuthal axes indicate the position of each point with respect to the flux tube axis and the colour scale shows the displacement of the flux surface at that point from its original position.
Motion away from and towards the axis of the flux tube are shown in red and blue, respectively.
The plots in the right-hand column show this information only for points at $\theta = 136.8^{\circ}$ and at $\theta = 316.8^{\circ}$, that is, points on opposite sides of the flux tube.
These angles are chosen to align with the direction of displacement of the flux tube to most clearly show the wave motion.
To aid the analysis of these plots the magnitude of velocity is also plotted in Figure \ref{fig:velmag} for a vertical slice through the centre of the domain.

Near the bottom of the domain (Figure \ref{fig:hslices}a, b), we see clear motion towards the axis on one side and away from it on the other, indicating a kink wave.
However, near the top (Figure \ref{fig:hslices}e, f), the motion on either side of the axis is in phase, either towards the axis on both sides or away from it on both sides.
This demonstrates the presence of a sausage wave.
In the middle of the domain (Figure \ref{fig:hslices}c, d), both waves are visible in different parts of the domain.
In this case, the kink wave is dominant close to the axis ($\lesssim 230$ km) and the sausage mode becomes more dominant further away.

This interpretation is consistent with the velocity magnitudes plotted in Figure \ref{fig:velmag}, which shows two distinct wave fronts.
One of these propagates almost isotropically and has reached the top of the domain, indicating that it is a fast mode.
The other wave front has not reached as great a height and is more closely confined to the magnetic field, and must therefore be the slow mode.
Since we can see in Figure \ref{fig:hslices} that only the sausage wave is visible at the top of the domain, this must correspond to the fast mode.
Similarly, the kink wave must correspond to the slow mode, since both are only seen close to the flux tube and in the lower half or so of the domain.

From this analysis we identify fast sausage waves and slow kink waves in our simulations without ambiguity.
This identification provides useful context to the rest of our analysis.

\subsection{Velocity components}
We select a field line at $r = 0.469$.
The changes in the velocities along this field line with time are plotted in the time-distance diagrams shown in Figure \ref{fig:vel-td} for the narrowest and widest drivers used in the simulations.
We also calculate the values of the fast speed ($v_f$), slow speed (also called the tube speed, $v_t$), sound speed ($c_s$) and Alfv\'en speed ($v_A$) along this field line (Equations \ref{eq:speeds}) and plot these for comparison.
These values are calculated thus:
\begin{subequations}
  \begin{align}
  c_s &= \sqrt{\frac{\gamma p}{\rho}} \\
  v_A &= \frac{B}{\sqrt{\mu_0 \rho}} \\
  v_f &= \sqrt{c_s^2 + v_A^2} \\
  v_t^{-2} &= \sqrt{c_s^{-2} + v_A^{-2}}
  \end{align}
  \label{eq:speeds}
\end{subequations}
where $p$ is the total (background plus perturbed) kinetic pressure.

Figure \ref{fig:vel-td} shows that for the narrowest driver, $0.15$ Mm ($1.67$ FWHM), the azimuthal velocity component is most dominant with an absolute value of $\sim 30$ ms$^{-1}$, compared to $\sim 16$ ms$^{-1}$ for both the parallel and perpendicular components.
For the widest driver, meanwhile, the absolute value of the azimuthal perturbation is lower ($\sim 16$ ms$^{-1}$), whereas the parallel and perpendicular perturbations have increased (to $\sim 60$ ms$^{-1}$ and $\sim 30$ ms$^{-1}$, respectively).
We see then that the azimuthal component is most dominant for a narrow driver, whereas a wider driver produces perturbations in which the parallel component is greatest.

\subsection{Flux contribution}

Time-distance diagrams for the contribution of each wave flux component are shown in Figure \ref{fig:flux-td}, again for the narrowest and widest drivers.
This contribution is expressed as the square of each flux component as a fraction of the total square flux, so that the sum of the three component contributions is equal to unity.

In the case of the widest driver, $0.35$ Mm ($3.89$ FWHM) we can see that the majority of the wave flux is contained in the parallel component.
This is particularly clear in the region above the height reached by slow magnetoacoustic and Alfv\'en waves, where some contribution comes from the perpendicular component but almost none can be seen in the azimuthal component.
The case for the narrowest driver is similar at these heights.
Below this, however, the relative importance of the perpendicular and azimuthal components is much greater compared to the widest driver.

Figure \ref{fig:fluxcompare-width} plots the percentage square wave flux, averaged over the full simulation run-time for each flux component and for each driver width.
This gives a broad indication of how dominant each component is over the simulation as a whole.
This comparison is plotted for each of three flux surfaces at different distances from the flux tube axis.
The parallel compenent varied between $\sim 50\%$ and $\sim 90\%$, the perpendicular component between $\sim 20\%$ and $\sim 10\%$, and the azimuthal component $\sim 35\%$ and $\sim 10\%$.
As spiral width increases, the influence of the parallel component increases, while the roles of the other components decrease.
Close to the axis of the flux tube, the contribution from the parallel component reaches almost $90\%$ for the widest driver, compared to around $65\%$ on the flux surface furthest from the axis.
The contributions of the perpendicular and azimuthal components are below $40\%$ for all flux surface radii and all driver widths, and both are almost universally much closer in value to each other than to the parallel component, apart from medium- and large distances from the flux tube axis for low-width drivers.

\section{Conclusions} \label{sec:conclusions}
The aim of this study was to investigate how the width of a photospheric spiral velocity driver affected the excited MHD waves in the lower solar atmosphere.
To achieve this, velocity profiles with a range of different widths between $0.15$ Mm ($1.67$ FWHM) and $0.35$ Mm ($3.89$ FWHM) were implemented to excite perturbations in a localised magnetic flux tube similar to one that might be found above a magnetic bright point (MBP).
The resulting perturbations were decomposed into parallel, perpendicular and azimuthal components and projected on to flux surfaces, and the corresponding wave energy fluxes calculated.
The relative contributions to the wave energy flux from these components were compared and evaluated.

First, these simulations do not include the transition region, where in reality waves might be reflected.
Whether or not such reflection takes place, and to what extent, will clearly have an impact on the amount of energy transmitted through the transition region into the corona.
Additionally, given the rapid expansion of the flux tube, it is likely that flux tubes which expand more gradually would display slightly different behaviour.
Slower expansion would lead to greater magnetic field strength near the top of the domain, assuming that other variables were kept the same.
This would change the plasma-$\beta$ and the fast, slow and Alfven speeds, all of which would affect the propagation of waves and thus may have implications for the amount of energy transfered to the corona.

The perpendicular component was found to have a minimal contribution for each spiral width, particularly on flux surfaces further from the centre of the domain.
Its contribution was greatest for the innermost flux surface and the narrowest driver, but even here it is quite small ($< 30\%$).
The azimuthal component behaves similarly, in that its contribution decreases for wider drivers, but unlike the perpendicular component, its contribution is greatest close to the centre of the domain.
Both components vary least on the largest flux surface.

The parallel component, on the other hand, has a significant flux contribution ($> 50\%$) for all drivers and all flux surfaces.
This contribution is greatest close to the flux tube axis and increases with driver width, reaching $\sim 90\%$ for the widest driver.

The effective excitation of the parallel wave component by these drivers is an important result, since we have shown it indicates the presence of a fast sausage mode.
It has been shown that this mode is ubiquitous in the quiet Sun and may carry enough energy to meet heating requirements in the chromosphere and low corona \citep{morton_observations_2012}.
Our results present a mechanism by which such waves could be excited by photospheric spiral velocity swirls consistent with observations.

\section*{Acknowledgements}
The authors are grateful to STFC (UK) for Grant No. ST/M000826/1.
VF, RE thank the Royal Society - Newton Mobility Grant for the support received.
VF also thanks the Newton Fund MAS/CONACyT Mobility Grants Program.
RE thanks the The Royal Society and the Chinese Academy of Sciences Presidents International Fellowship Initiative, Grant No. 2016VMA045 for support received.
The authors also thank the NumPy, SciPy \citep{van_der_walt_numpy_2011}, Matplotlib \citep{hunter_matplotlib:_2007}, MayaVi2 \citep{ramachandran_mayavi:_2011}, yt \citep{turk_yt:_2011} and Astropy \citep{the_astropy_collaboration_astropy:_2013} Python projects for providing the tools to analyse and visualise the data.

\bibliographystyle{apalike}
\bibliography{refs}

\begin{thebibliography}{}

\bibitem[Bogdan and Judge, 2006]{bogdan_observational_2006}
Bogdan, T. and Judge, P. (2006).
\newblock Observational aspects of sunspot oscillations.
\newblock {\em Philosophical Transactions of the Royal Society A: Mathematical,
  Physical and Engineering Sciences}, 364(1839):313--331.

\bibitem[Bogdan et~al., 2003]{bogdan_waves_2003}
Bogdan, T.~J., Hansteen, M.~C.~V., McMurry, A., Rosenthal, C.~S., Johnson, M.,
  Petty-Powell, S., Zita, E.~J., Stein, R.~F., McIntosh, S.~W., and Nordlund,
  {\r a}. (2003).
\newblock Waves in the {Magnetized} {Solar} {Atmosphere}. {II}. {Waves} from
  {Localized} {Sources} in {Magnetic} {Flux} {Concentrations}.
\newblock {\em The Astrophysical Journal}, 599(1):626--660.

\bibitem[Bonet et~al., 2010]{bonet_sunrise/imax_2010}
Bonet, J.~A., M{\'a}rquez, I., Almeida, J.~S., Palacios, J., Pillet, V.~M.,
  Solanki, S.~K., del Toro~Iniesta, J.~C., Domingo, V., Berkefeld, T., Schmidt,
  W., Gandorfer, A., Barthol, P., and Kn{\"o}lker, M. (2010).
\newblock {SUNRISE}/{IMaX} {OBSERVATIONS} {OF} {CONVECTIVELY} {DRIVEN} {VORTEX}
  {FLOWS} {IN} {THE} {SUN}.
\newblock {\em The Astrophysical Journal}, 723(2):L139--L143.

\bibitem[Bonet et~al., 2008]{bonet_convectively_2008}
Bonet, J.~A., M{\'a}rquez, I., S{\'a}nchez~Almeida, J., Cabello, I., and
  Domingo, V. (2008).
\newblock Convectively {Driven} {Vortex} {Flows} in the {Sun}.
\newblock {\em The Astrophysical Journal}, 687(2):L131--L134.

\bibitem[De~Moortel et~al., 2016]{de_moortel_transverse_2016}
De~Moortel, I., Pascoe, D.~J., Wright, A.~N., and Hood, A.~W. (2016).
\newblock Transverse, propagating velocity perturbations in solar coronal
  loops.
\newblock {\em Plasma Physics and Controlled Fusion}, 58(1):014001.

\bibitem[Deinzer, 1965]{deinzer_magneto-hydrostatic_1965}
Deinzer, W. (1965).
\newblock On the {Magneto}-{Hydrostatic} {Theory} of {Sunspots}.
\newblock {\em The Astrophysical Journal}, 141:548.

\bibitem[Dorotovi{\v c} et~al., 2014]{dorotovic_standing_2014}
Dorotovi{\v c}, I., Erd{\'e}lyi, R., Freij, N., Karlovsk{\'y}, V., and
  M{\'a}rquez, I. (2014).
\newblock Standing sausage waves in photospheric magnetic waveguides.
\newblock {\em Astronomy \& Astrophysics}, 563:A12.

\bibitem[Fedun et~al., 2011a]{fedun_numerical_2011}
Fedun, V., Shelyag, S., and Erd{\'e}lyi, R. (2011a).
\newblock {NUMERICAL} {MODELING} {OF} {FOOTPOINT}-{DRIVEN} {MAGNETO}-{ACOUSTIC}
  {WAVE} {PROPAGATION} {IN} {A} {LOCALIZED} {SOLAR} {FLUX} {TUBE}.
\newblock {\em The Astrophysical Journal}, 727(1):17.

\bibitem[Fedun et~al., 2011b]{fedun_mhd_2011}
Fedun, V., Shelyag, S., Verth, G., Mathioudakis, M., and Erd{\'e}lyi, R.
  (2011b).
\newblock {MHD} waves generated by high-frequency photospheric vortex motions.
\newblock {\em Annales Geophysicae}, 29(6):1029--1035.

\bibitem[Freij et~al., 2014]{freij_detection_2014}
Freij, N., Scullion, E.~M., Nelson, C.~J., Mumford, S., Wedemeyer, S., and
  Erd{\'e}lyi, R. (2014).
\newblock {THE} {DETECTION} {OF} {UPWARDLY} {PROPAGATING} {WAVES} {CHANNELING}
  {ENERGY} {FROM} {THE} {CHROMOSPHERE} {TO} {THE} {LOW} {CORONA}.
\newblock {\em The Astrophysical Journal}, 791(1):61.

\bibitem[Gent et~al., 2013]{gent_magnetohydrostatic_2013}
Gent, F.~A., Fedun, V., Mumford, S.~J., and Erdelyi, R. (2013).
\newblock Magnetohydrostatic equilibrium - {I}. {Three}-dimensional open
  magnetic flux tube in the stratified solar atmosphere.
\newblock {\em Monthly Notices of the Royal Astronomical Society},
  435(1):689--697.

\bibitem[Hunter, 2007]{hunter_matplotlib:_2007}
Hunter, J.~D. (2007).
\newblock Matplotlib: {A} 2d {Graphics} {Environment}.
\newblock {\em Computing in Science \& Engineering}, 9(3):90--95.

\bibitem[Jess et~al., 2009]{jess_alfven_2009}
Jess, D.~B., Mathioudakis, M., Erdelyi, R., Crockett, P.~J., Keenan, F.~P., and
  Christian, D.~J. (2009).
\newblock Alfven {Waves} in the {Lower} {Solar} {Atmosphere}.
\newblock {\em Science}, 323(5921):1582--1585.

\bibitem[Kitiashvili et~al., 2013]{kitiashvili_ubiquitous_2013}
Kitiashvili, I.~N., Kosovichev, A.~G., Lele, S.~K., Mansour, N.~N., and Wray,
  A.~A. (2013).
\newblock {UBIQUITOUS} {SOLAR} {ERUPTIONS} {DRIVEN} {BY} {MAGNETIZED} {VORTEX}
  {TUBES}.
\newblock {\em The Astrophysical Journal}, 770(1):37.

\bibitem[Leroy, 1985]{leroy_derivation_1985}
Leroy, B. (1985).
\newblock On the derivation of the energy flux of linear magnetohydrodynamic
  waves.
\newblock {\em Geophysical \& Astrophysical Fluid Dynamics}, 32(2):123--133.

\bibitem[Mathioudakis et~al., 2013]{mathioudakis_alfven_2013}
Mathioudakis, M., Jess, D.~B., and Erd{\'e}lyi, R. (2013).
\newblock Alfv{\'e}n {Waves} in the {Solar} {Atmosphere}: {From} {Theory} to
  {Observations}.
\newblock {\em Space Science Reviews}, 175(1-4):1--27.

\bibitem[McIntosh et~al., 2011]{mcintosh_alfvenic_2011}
McIntosh, S.~W., De~Pontieu, B., Carlsson, M., Hansteen, V., Boerner, P., and
  Goossens, M. (2011).
\newblock Alfv{\'e}nic waves with sufficient energy to power the quiet solar
  corona and fast solar wind.
\newblock {\em Nature}, 475(7357):477--480.

\bibitem[Morton et~al., 2012]{morton_observations_2012}
Morton, R.~J., Verth, G., Jess, D.~B., Kuridze, D., Ruderman, M.~S.,
  Mathioudakis, M., and Erd{\'e}lyi, R. (2012).
\newblock Observations of ubiquitous compressive waves in the
  {Sun}{\textquoteright}s chromosphere.
\newblock {\em Nature Communications}, 3:1315.

\bibitem[Mumford, 2016]{mumford_2016_48888}
Mumford, S.~J. (2016).
\newblock {\em {Simulations of Magnetohydrodynamic Waves Driven by Photospheric
  Motions}}.
\newblock PhD thesis, The University of Sheffield.

\bibitem[Mumford et~al., 2015]{mumford_generation_2015}
Mumford, S.~J., Fedun, V., and Erd{\'e}lyi, R. (2015).
\newblock {GENERATION} {OF} {MAGNETOHYDRODYNAMIC} {WAVES} {IN} {LOW} {SOLAR}
  {ATMOSPHERIC} {FLUX} {TUBES} {BY} {PHOTOSPHERIC} {MOTIONS}.
\newblock {\em The Astrophysical Journal}, 799(1):6.

\bibitem[Nakariakov and Verwichte, 2005]{nakariakov_coronal_2005}
Nakariakov, V.~M. and Verwichte, E. (2005).
\newblock Coronal {Waves} and {Oscillations}.
\newblock {\em Living Reviews in Solar Physics}, 2.

\bibitem[Ramachandran and Varoquaux, 2011]{ramachandran_mayavi:_2011}
Ramachandran, P. and Varoquaux, G. (2011).
\newblock Mayavi: 3d {Visualization} of {Scientific} {Data}.
\newblock {\em Computing in Science \& Engineering}, 13(2):40--51.

\bibitem[S{\'a}nchez~Almeida et~al., 2004]{sanchez_almeida_bright_2004}
S{\'a}nchez~Almeida, J., M{\'a}rquez, I., Bonet, J.~A.,
  Dom{\'i}nguez~Cerde{\~n}a, I., and Muller, R. (2004).
\newblock Bright {Points} in the {Internetwork} {Quiet} {Sun}.
\newblock {\em The Astrophysical Journal}, 609(2):L91--L94.

\bibitem[Schluter and Temesvary, 1958]{schluter_internal_1958}
Schluter, A. and Temesvary, S. (1958).
\newblock The {Internal} {Constitution} of {Sunspots}.
\newblock In {\em Electromagnetic {Phenomena} in {Cosmical} {Physics}}.

\bibitem[Sch{\"u}ssler and Rempel, 2005]{schussler_dynamical_2005}
Sch{\"u}ssler, M. and Rempel, M. (2005).
\newblock The dynamical disconnection of sunspots from their magnetic roots.
\newblock {\em Astronomy and Astrophysics}, 441(1):337--346.

\bibitem[{Sekse} et~al., 2013]{2013ApJ...769...44S}
{Sekse}, D.~H., {Rouppe van der Voort}, L., {De Pontieu}, B., and {Scullion},
  E. (2013).
\newblock {Interplay of Three Kinds of Motion in the Disk Counterpart of Type
  II Spicules: Upflow, Transversal, and Torsional Motions}.
\newblock {\em \apj}, 769:44.

\bibitem[Shelyag et~al., 2008]{shelyag_magnetohydrodynamic_2008}
Shelyag, S., Fedun, V., and Erd{\'e}lyi, R. (2008).
\newblock Magnetohydrodynamic code for gravitationally-stratified media.
\newblock {\em Astronomy and Astrophysics}, 486(2):655--662.

\bibitem[Shelyag et~al., 2011]{shelyag_photospheric_2011}
Shelyag, S., Fedun, V., Keenan, F.~P., Erd{\'e}lyi, R., and Mathioudakis, M.
  (2011).
\newblock Photospheric magnetic vortex structures.
\newblock {\em Annales Geophysicae}, 29(5):883--887.

\bibitem[{The Astropy Collaboration} et~al.,
  2013]{the_astropy_collaboration_astropy:_2013}
{The Astropy Collaboration}, Robitaille, T.~P., Tollerud, E.~J., Greenfield,
  P., Droettboom, M., Bray, E., Aldcroft, T., Davis, M., Ginsburg, A.,
  Price-Whelan, A.~M., Kerzendorf, W.~E., Conley, A., Crighton, N., Barbary,
  K., Muna, D., Ferguson, H., Grollier, F., Parikh, M.~M., Nair, P.~H.,
  G{\"u}nther, H.~M., Deil, C., Woillez, J., Conseil, S., Kramer, R., Turner,
  J. E.~H., Singer, L., Fox, R., Weaver, B.~A., Zabalza, V., Edwards, Z.~I.,
  Azalee~Bostroem, K., Burke, D.~J., Casey, A.~R., Crawford, S.~M., Dencheva,
  N., Ely, J., Jenness, T., Labrie, K., Lim, P.~L., Pierfederici, F., Pontzen,
  A., Ptak, A., Refsdal, B., Servillat, M., and Streicher, O. (2013).
\newblock Astropy: {A} community {Python} package for astronomy.
\newblock {\em Astronomy \& Astrophysics}, 558:A33.

\bibitem[T{\'o}th, 1996]{toth_general_1996}
T{\'o}th, G. (1996).
\newblock A {General} {Code} for {Modeling} {MHD} {Flows} on {Parallel}
  {Computers}: {Versatile} {Advection} {Code}.
\newblock {\em Astrophysical Letters and Communications}, 34:245.

\bibitem[Turk et~al., 2011]{turk_yt:_2011}
Turk, M.~J., Smith, B.~D., Oishi, J.~S., Skory, S., Skillman, S.~W., Abel, T.,
  and Norman, M.~L. (2011).
\newblock yt: {A} {MULTI}-{CODE} {ANALYSIS} {TOOLKIT} {FOR} {ASTROPHYSICAL}
  {SIMULATION} {DATA}.
\newblock {\em The Astrophysical Journal Supplement Series}, 192(1):9.

\bibitem[van~der Walt et~al., 2011]{van_der_walt_numpy_2011}
van~der Walt, S., Colbert, S.~C., and Varoquaux, G. (2011).
\newblock The {NumPy} {Array}: {A} {Structure} for {Efficient} {Numerical}
  {Computation}.
\newblock {\em Computing in Science \& Engineering}, 13(2):22--30.

\bibitem[Vernazza et~al., 1981]{vernazza_structure_1981}
Vernazza, J.~E., Avrett, E.~H., and Loeser, R. (1981).
\newblock Structure of the solar chromosphere. {III} - {Models} of the {EUV}
  brightness components of the quiet-sun.
\newblock {\em The Astrophysical Journal Supplement Series}, 45:635.

\bibitem[Vigeesh et~al., 2012]{vigeesh_three-dimensional_2012}
Vigeesh, G., Fedun, V., Hasan, S.~S., and Erd{\'e}lyi, R. (2012).
\newblock {THREE}-{DIMENSIONAL} {SIMULATIONS} {OF} {MAGNETOHYDRODYNAMIC}
  {WAVES} {IN} {MAGNETIZED} {SOLAR} {ATMOSPHERE}.
\newblock {\em The Astrophysical Journal}, 755(1):18.

\bibitem[Wang, 2011]{wang_standing_2011}
Wang, T. (2011).
\newblock Standing {Slow}-{Mode} {Waves} in {Hot} {Coronal} {Loops}:
  {Observations}, {Modeling}, and {Coronal} {Seismology}.
\newblock {\em Space Science Reviews}, 158(2-4):397--419.

\bibitem[Wedemeyer-B{\"o}hm and Rouppe van~der Voort,
  2009]{wedemeyer-bohm_small-scale_2009}
Wedemeyer-B{\"o}hm, S. and Rouppe van~der Voort, L. (2009).
\newblock Small-scale swirl events in the quiet {Sun} chromosphere.
\newblock {\em Astronomy and Astrophysics}, 507(1):L9--L12.

\bibitem[Wedemeyer-B{\"o}hm et~al., 2012]{wedemeyer-bohm_magnetic_2012}
Wedemeyer-B{\"o}hm, S., Scullion, E., Steiner, O., van~der Voort, L.~R., de~la
  Cruz~Rodriguez, J., Fedun, V., and Erd{\'e}lyi, R. (2012).
\newblock Magnetic tornadoes as energy channels into the solar corona.
\newblock {\em Nature}, 486(7404):505--508.

\bibitem[Zaqarashvili and Erd{\'e}lyi, 2009]{zaqarashvili_oscillations_2009}
Zaqarashvili, T.~V. and Erd{\'e}lyi, R. (2009).
\newblock Oscillations and {Waves} in {Solar} {Spicules}.
\newblock {\em Space Science Reviews}, 149(1-4):355--388.

\end{thebibliography}

\label{lastpage}
\end{document}